\documentclass[amsmath,twocolumn]{aastex702}

\accepted{\today}

\submitjournal{APJ}

\begin{document}

\title{
  Uncovering the Origin of Slow Rotators among Intermediate-Mass Stars in the Star-Forming Cluster Trumpler\,14
       }

\author[0000-0001-7470-9192]{Khushboo K. Rao}
\affiliation{Institute of Astronomy, National Central University, 320317 Taoyuan, Taiwan}
\email[show]{khushboo@astro.ncu.edu.tw}  

\author[0000-0003-0262-272X]{Wen-Ping Chen}
\affiliation{Institute of Astronomy, National Central University, 320317 Taoyuan, Taiwan}
\affiliation{Department of Physics,  National Central University, 320317 Taoyuan, Taiwan}
\email{wchen@astro.ncu.edu.tw}  

\author[0009-0001-2922-2315]{Hsiang-Yu Chen}
\affiliation{Institute of Astronomy, National Central University, 320317 Taoyuan, Taiwan}
\email{xy.chen429@gmail.com}  

\author[0000-0002-8532-4025]{Kristen Dage}
\affiliation{International Centre for Radio Astronomy Research -- Curtin University, GPO Box U1987, Perth, WA 6845, Australia}
\email{kristen.dage@curtin.edu.au}  

\author[0000-0003-1252-9916]{Stella S. R. Offner}
\affiliation{University of Texas, Austin, Texas, USA}
\email{soffner@astro.as.utexas.edu}  


\begin{abstract}

Intermediate-mass (about 1.5--8~M$_{\odot}$) stars exhibit a wide range of rotation rates and have gained attention for their roles in the extended main-sequence turnoffs (eMSTO) seen in some young and intermediate-age clusters (ages $< 2$~Gyr). Although rapid rotation is expected due to their radiative envelopes, the presence of slow rotators remains puzzling. In this study, we examine disk and X-ray signatures among the intermediate-mass members of the star-forming cluster Trumpler\,14. Of the 118 intermediate-mass members, 24 are considered disk-bearing on the basis of their infrared spectral indices derived from spectral energy distributions.  The majority of these reside outside the heavily irradiated cluster core and are all slow rotators ($\lessapprox 100$~km~s$^{-1}$) except for three fast-rotating Class~III objects.
Additionally, in the same sample of 118, 37 are X-ray sources, and 22 having available $v \sin i$ data show a systematically slower rotation than X-ray quiet sources. Our findings suggest that, while young stellar disks in intermediate-mass stars contribute to early spin-down, high-energy processes traced by X-ray emission during this phase could be an additional channel for angular-momentum loss in these stars. 
\end{abstract}


\section{Introduction} \label{sec:intro}
Clusters younger than 2~Gyr serve as valuable testbeds for studying star formation and early evolution. In particular, the intermediate-mass (1.5--8~M$_\odot$) members, with a distributed range of projected rotational velocities ($v \sin i$) ranging from $< 10$~km~s$^{-1}$ to 100--400~km~s$^{-1}$ \citep{Bastian2018MNRAS.480.3739B, Milone2018MNRAS.477.2640M, Kamann2020MNRAS.492.2177K, Kamann2023MNRAS.518.1505K, Kamann2025MNRAS.542.2768K, Cordoni2024MNRAS.532.1547C}, manifest the role of rotation \citep{Bastian2009MNRAS.398L..11B} in broadening the upper main sequence (MS) and the main-sequence turnoff (MSTO) regions in the color-magnitude diagrams (CMDs) of these systems. 

Rapid rotation modifies stellar structure through rotational deformation and induces gravity darkening \citep{Espinosa2011A&A...533A..43E}, while the observed photometric properties further depend on the viewing inclination \citep{Georgy2014A&A...566A..21G, Groh2019A&A...627A..24G, Nguyen2022A&A...665A.126N}. Collectively for a star cluster, stars with different rotation rates occupy different locations in the CMD, with rapidly rotating stars appearing redder than their slowly rotating counterparts. This distribution broadens the upper MS and MSTOs, producing the extended main sequence (eMS) and extended main-sequence turnoff (eMSTO) features observed in numerous young and intermediate-age clusters \citep{Bastian2009MNRAS.398L..11B, Mackey2007MNRAS.379..151M, Milone2009A&A...497..755M}, some of which even show clear bifurcated MSTOs \citep{Mackey2007MNRAS.379..151M, Milone2016MNRAS.458.4368M, Marino2018AJ....156..116M}.

Fast rotation is understood when pre–main-sequence (PMS) objects contract onto the zero-age MS. Unlike low-mass stars, which experience subsequent magnetic braking, intermediate- or high-mass stars, owing to their radiative envelopes and thus generally not magnetized, remain fast-rotating.  
The existence of a substantial population of very slow rotators, even after the inclination effect is considered, is hence puzzling.   \citet{Rao2026ApJ..1002..103R}, using a comprehensive sample of Milky Way clusters from 10~Myr to 1~Gyr age, reported that $\approx 8\%$ of the intermediate-mass stars have $v \sin i< 30$~km~s$^{-1}$, and $\approx 40\%$ have $v \sin i < 100$~km~s$^{-1}$.  In the LMC, NGC\,1846 of $\sim 1.5$~Gyr age and NGC\,1831 of $\sim 800$~Myr age have 40--63\% members as slow rotators \citep{Kamann2020MNRAS.492.2177K, Correnti2021MNRAS.504..155C}, and in NGC\,1783 at an age of $\sim 1.5$~Gyr, 18\% of its MSTO members rotate very slowly \citep[$v \sin i < 50$~km~s$^{-1}$;][]{Leanza2025A&A...698A..27L}.

Several mechanisms have been proposed to account for slow rotation, including tidal synchronization in binaries \citep{DAntona2015MNRAS4532637}, stellar mergers \citep{Wang2022NatAs...6..480W}, and PMS star-disk interactions \citep{Bastian2020MNRAS.495.1978B}. However, recent studies indicate that the prevalence of slow rotators cannot be fully explained by tidal synchronization in close binaries alone \citep{He2022ApJ...938...42H, Maurya2024MNRAS.532.1212M}. Additionally, a merging event may actually spin up the merged product instead \citep{Bastian2025A&A...700A.241B}, with its subsequent rotational velocity dependent on several factors \citep{Schneider2025arXiv250918421S}. Therefore, the key diagnosis is the angular momentum evolution of intermediate-mass stars during the PMS phase.

For solar-type and lower-mass PMS stars, the angular momentum is regulated by, e.g., magnetized stellar winds, star-disk magnetic coupling, episodic magnetospheric ejections, jets, or disk winds, each of which counterbalances the angular momentum injected by accreted material \citep[see][and references therein]{Bouvier2014prpl.conf..433B, Bastian2020MNRAS.495.1978B}. Notably, the star-disk coupling plays a pivotal role prior to the T~Tauri phase, as a fully convective protostar anchors the magnetic fields onto the circumstellar disk, leading to co-rotation of the star with the disk, thereby spinning down the central collapsing body that otherwise would have rotated near breakup \citep{Shu1988ApJ...328L..19S}. Are any of these applicable in intermediate-mass stars during early evolution, resulting in substantial spin-down before reaching the MS?  
Disk-bearing intermediate-mass PMS objects are called Herbig Ae/Be stars. They share similar observables as solar-type PMS objects, namely, T~Tauri stars, such as infrared excess due to thermal emission from circumstellar dust, emission spectra, and X-ray emission arising from chromospheric/coronal activities.
\citet{Bu2025ApJ...979...29B} showed that the Herbig Ae/Be members in the young cluster NGC\,2264 are preferentially slow rotators compared to their diskless counterparts.

Newly formed objects also emit copious X rays throughout the mass spectrum from protostars to massive stars \citep{preibisch05a,preibisch05c}. For example, the Chandra Carina Complex project \citep{Townsley2011ApJS..194....1T} detected more than 10000 X-ray sources. T~Tauri stars with deep convective envelopes are magnetized. As young objects contract toward the MS, the extended chromosphere/corona produces prominent emission-line spectra and X-rays, more so than their evolved MS counterparts \citep{Grosso1997Natur.387...56G}. Seen in the Sun, the magnetic field lines may reconnect, energizing plasma to cause sudden brightening events detected from gamma rays to radio wavelengths.  Such flares are also observed in red or even brown dwarfs, especially among the young ones that have not been spun down via magnetic braking.  If the field lines, rather than looping back to the stellar surface itself, are anchored to a circumstellar disk, a stellar companion, or an orbiting planet, the reconnection may occur on a large scale in volume and in amount of energy, as observed in W~UMa or RS~CVn binaries \citep{Fleming1989AJ.....98..692F}. Some red dwarfs may be prone to flares because of a close-in magnetized planet \citep{Lin2022AJ....163..164L}.  

However, massive OB stars with convective cores, though a fossil field may exist, are not believed to be magnetically active and thus owe their X-ray emission to powerful colliding winds between usually equally massive companions \citep{Rauw2016AdSpR..58..761R}. 
The origin of X rays from intermediate-mass stars is not clear. Once on the MS, they are not expected to have a surface magnetic field; therefore, the energetic emission is generally attributable either to powerful wind shocks, as in an OB pair, or to an unresolved T~Tauri companion. However, in the PMS phase, intermediate-mass stars, namely Herbig Ae/Be stars, may still be convective, albeit with comparatively short Hayashi-track lifetimes, and can therefore produce X rays through residual magnetic fields and radiative shear \citep{Tout1995MNRAS.272..528T, Hubrig2009A&A...502..283H}.

\begin{figure*}
\begin{center}
    \includegraphics[width=0.32\linewidth]{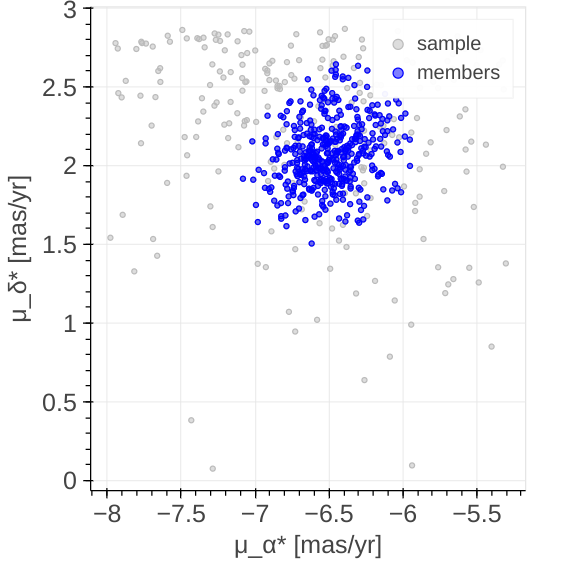}
    \includegraphics[width=0.32\linewidth]{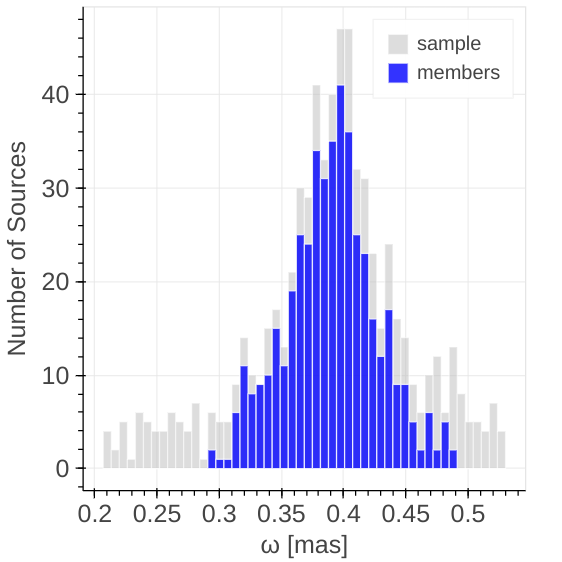}
    \includegraphics[width=0.32\linewidth]{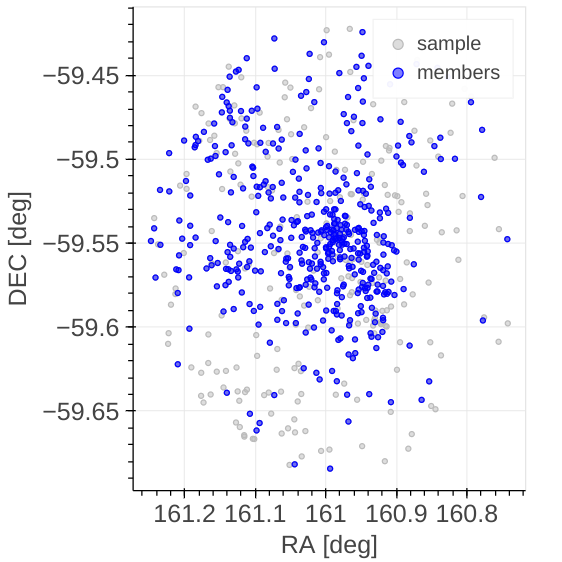}
    \caption{Proper motion, parallax, and spatial distribution of member candidates (blue dots) against field stars (grey dots) toward the Trumpler\,14 open cluster. }
    \label{fig:membership}
\end{center}
\end{figure*}

The Carina star-forming region, rich in young clusters with OB and intermediate-mass stars \citep{Nunez2021AJ....162..153N, Goppl2025A&A...695A..48G, 1988PASP..100.1134B}, is suitable for investigating angular momentum regulation during the PMS phase. In this work, we carry out a multiwavelength study of the intermediate-mass population of the rich open cluster Trumpler\,14 (1--10 Myr). As a part of the Carina complex, the cluster has been included in several X-ray-based studies \citep{Sanchawala2007ApJ...656..462S, Gagne2011ApJS..194....5G, Damiani2017A&A...603A..81D}. An early study by \citet{damiani94} reported that the X-ray luminosities of Herbig Ae/Be stars were not correlated with $v \sin i$. Here we revisit this subject, namely the interplay among X-ray activity, the presence of circumstellar dusty disks, and angular-momentum regulation of the intermediate-mass PMS population in the Trumpler\,14 open cluster. 

%
\section{Data and Membership} \label{sec:data}
We used the ML-MOC algorithm \citep{Agarwal2021} on \textit{Gaia} DR3 data \citep{Gaia2023A&A...674A...1G} to identify cluster members. ML-MOC employs a machine-learning approach that integrates the k-Nearest Neighbors (kNN) and Gaussian Mixture Model (GMM) algorithms, leveraging proper motions and parallaxes to determine cluster membership. Initially, kNN is applied to remove the obvious field stars, and then GMM is applied to obtain a final list of members with their membership probabilities. In the GMM, the data are assumed to be composed of two Gaussians, one belonging to the cluster and the other to the field. We identified a total of 452 members within the 8~arcmin field of Trumpler\,14.  The proper motions, parallaxes, and spatial distribution of the cluster members against field stars are shown in Fig.~\ref{fig:membership}. The estimated radius for the cluster is 5~arcmin, beyond which the surface density of members appears to merge with that of the field stars. Of these 452 cluster members, 342 sources are located within a 5~arcmin radius and are used for further analysis.  The precise size of the cluster is not critical in our work.

\begin{figure}
\begin{center}
    \includegraphics[width=0.97\linewidth]{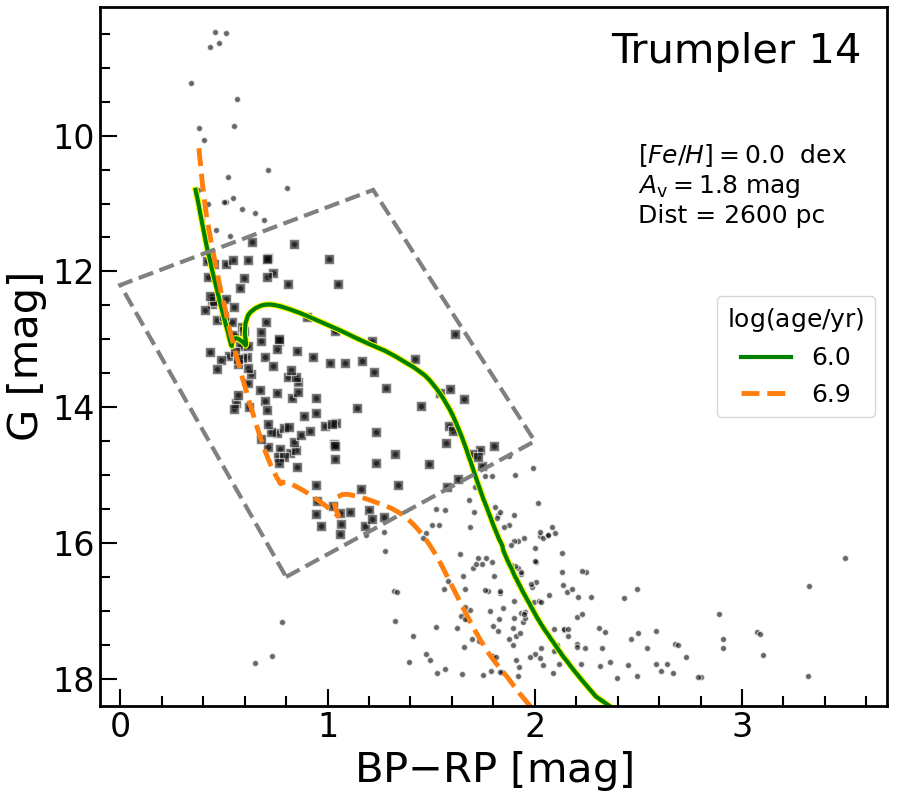}
    \caption{The CMD of Trumpler\,14 plotted with non-rotating PARSEC 2.0 isochrones of $\log({\rm Age/yr})  =  6.0$ and 6.9, a distance of $2600$~pc, $[Fe/H] = 0.0$~dex, and $A_{\rm V} = 1.8$~mag. The MS turn-on sources selected for spectral energy distribution analysis are shown as black squares inside the grey dashed box.}
    \label{fig:cmd}
\end{center}
\end{figure}

The \textit{Gaia} CMD of the 342 cluster members is shown in Fig.~\ref{fig:cmd} with fitted non-rotating PARSEC 2.0 isochrones \citep{Nguyen2022A&A...665A.126N}. To fit the isochrones, initial parameters such as age, $A{_{\rm V}}$, metallicity, and distance are taken from the literature \citep[e.g.,][]{DeGioia2001ApJ...549..578D, Bragaglia2022A&A...659A.200B, Itrich2024A&A...685A.100I}. These parameters are then adjusted to visually best fit the isochrones to the observed CMD. The final parameters of the cluster are: $\log({\rm Age/yr}) = 6.0, 6.9$, distance = 2600~pc, metallicity $([M/H]) = 0.0$~dex, and visual extinction of $(A{_{\rm V}})= 1.8$~mag.  These cluster parameters are consistent with those reported in the literature \citep[e.g.,][]{Itrich2024A&A...685A.100I, DeGioia2001ApJ...549..578D}. 

\begin{figure*}
\begin{center}
    \includegraphics[width=0.99\textwidth]{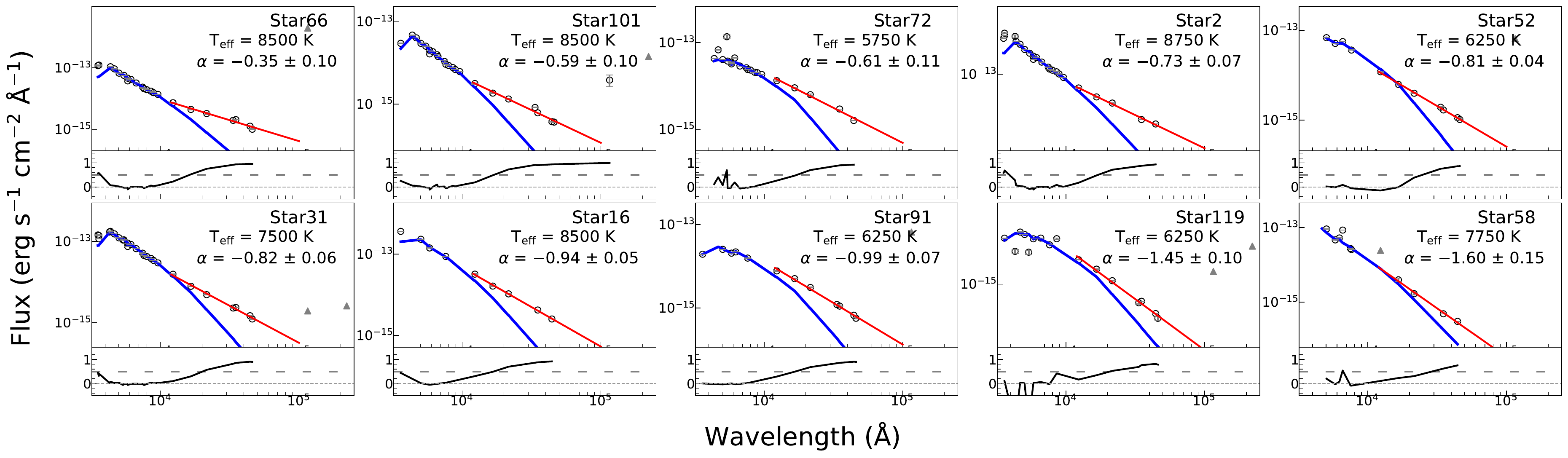}
    \includegraphics[width=0.99\textwidth]{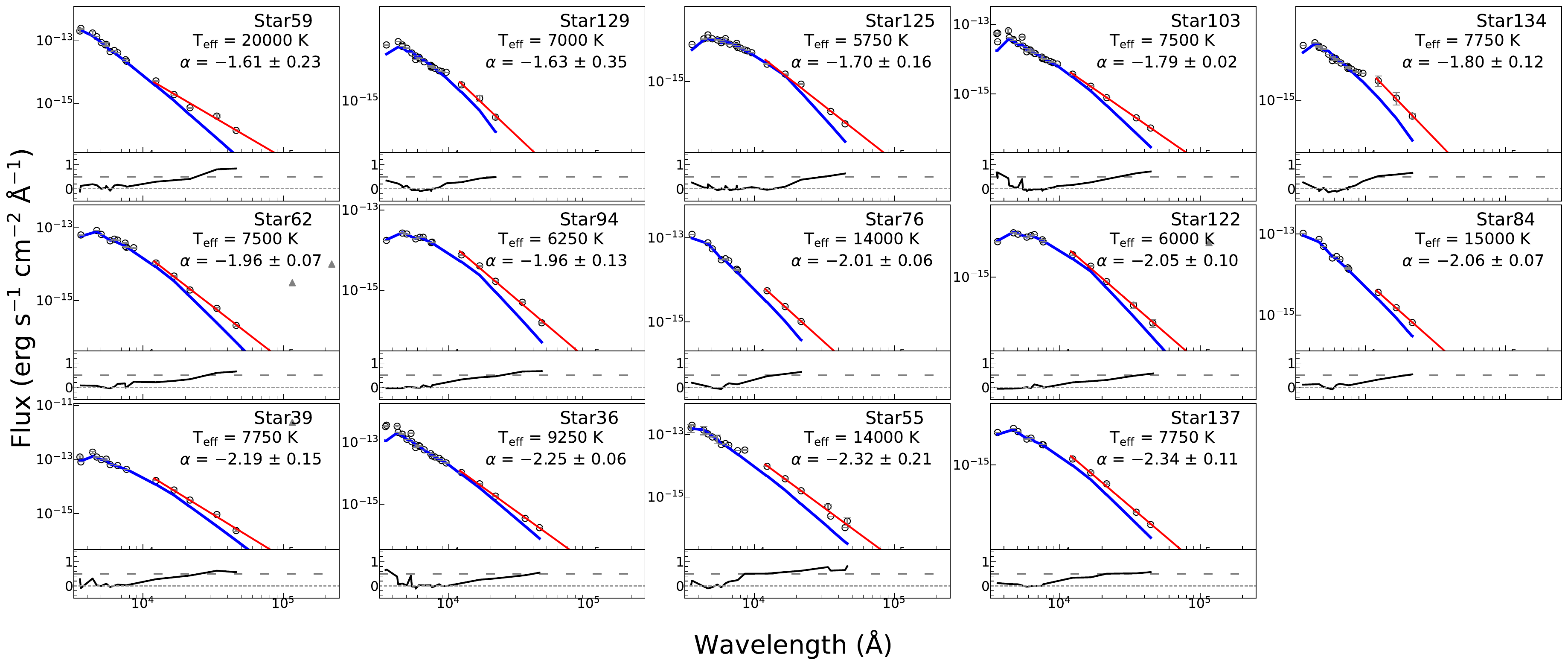}
    \caption{Objects showing prominent infrared excess of Class~II (upper panel) and Class~III (lower panel) with the infrared spectral index $\alpha > -2.34$ in ascending order. Circle symbols represent observed data, with the grey triangles marking upper limits. In each case, the blue line depicts the fitted Kurucz atmospheric model, and the red line shows the best-fit $\alpha$ index to the IR data, spanning from the 2MASS $J$ band to the WISE $W2$ band or Spitzer/\textit{IRAC} $I2$ whenever available. In the lower panel of each SED, the grey dotted and dashed lines mark the 0 and 0.5 fractional flux residual levels, respectively. The black solid lines represent the observed fractional flux residuals.}
    \label{fig:IR_excess_SED}
\end{center}
\end{figure*}

The $v \sin i$ values for stars are generally estimated by broadening of spectral lines. However, equatorial velocities ($v_{\rm eq}$) require additional knowledge of the inclination angle, which is not readily available. Therefore, in this work, the $v \sin i$ measurements are used as a proxy for $v_{\rm eq}$ and are taken from various spectroscopic surveys  \citep{Damiani2017A&A...603A..81D, Hanes2018AJ....155..190H, Abdurro2022ApJS..259...35A, Randich2022A&A...666A.121R, Sprague2022AJ....163..152S, Hourihane2023A&A...676A.129H, Berlanas2025A&A...695A.248B}. Details of the $v \sin i $ measurements are presented in Appendix~\ref{sec:vsini}.

The \textit{Gaia} ESO DR5.1  \citep{Hourihane2023A&A...676A.129H} provide radial velocity measurements for 118 cluster members, more than other spectroscopic surveys available for the cluster. The mean radial velocity for these 118 members is $-7.66$~km~s$^{-1}$, which is in close agreement with the value reported by \citet{Dias2002A&A...389..871D}. Of these 118 sources, 90 fall within the $-7.66 \pm 20$~km~s$^{-1}$ range. As an example, a spectroscopic binary (\#Star1; \textit{Gaia} DR3 ID = 5350363905936139904) has a radial velocity of 63.75~km~s$^{-1}$ from \textit{Gaia} DR3 and $-43.66$~km~s$^{-1}$ \textit{Gaia} ESO DR5.1 data. This star is included as a cluster member.  Therefore, in the absence of information on binarity, we do not exclude membership based on radial velocity measurements. In other words, cluster membership is determined purely based on proper motions and parallaxes.
%
\section{Results}
\subsection{Estimation of the infrared spectral index}
%
\begin{figure*}
\begin{center}
    \includegraphics[width=0.48\textwidth]{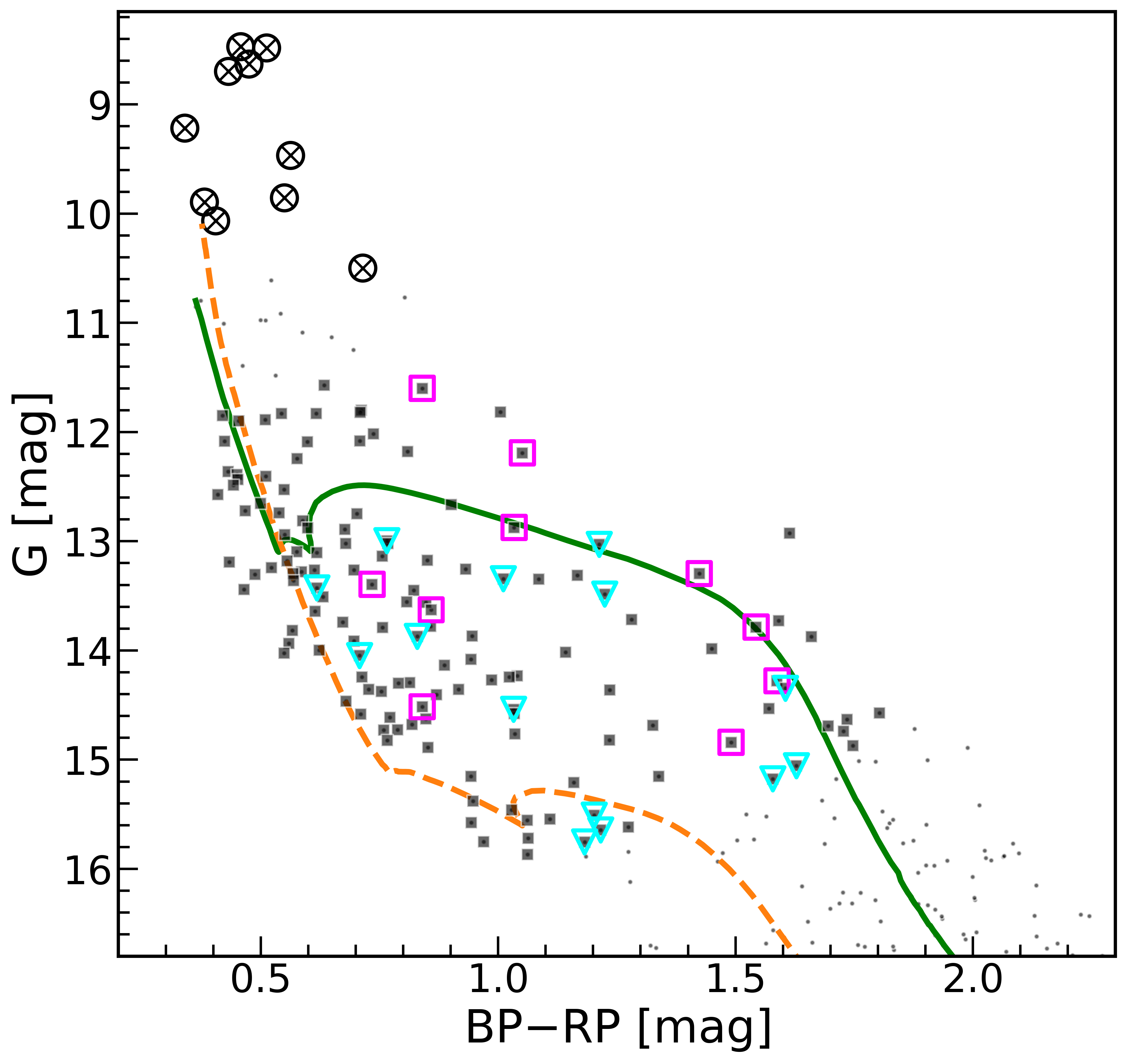}
    \includegraphics[width=0.48\textwidth]{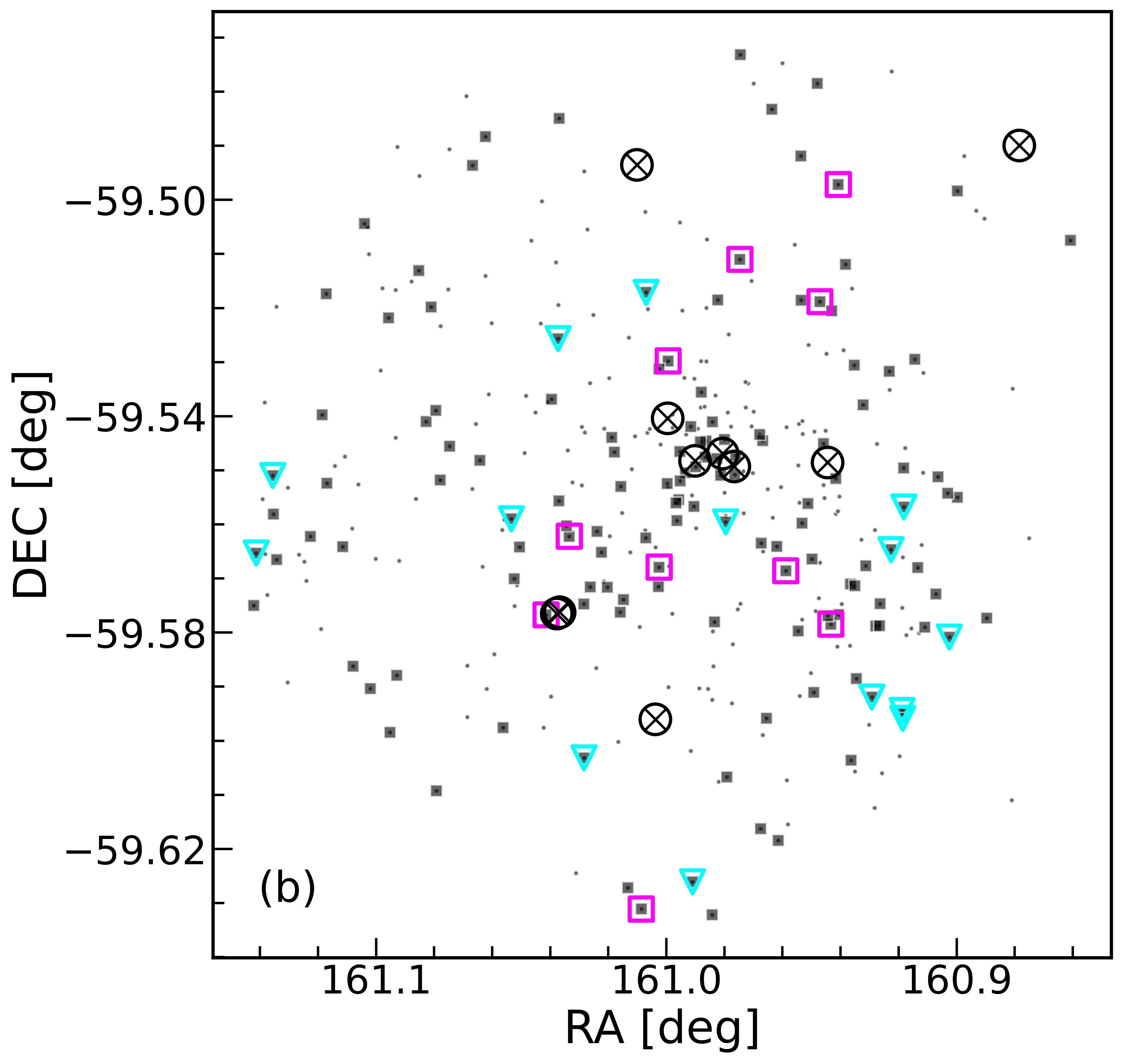}
    \caption{(a)~The CMD as in Fig~\ref{fig:cmd} but is separately marked with massive O-type members in big black circles,  Class~II candidates in magenta squares, and Class~III candidates in cyan triangles. (b)~The spatial distribution of the objects shown in (a) with the same symbols.}
    \label{fig:CMD_RADEC}
\end{center}
\end{figure*}

A total of 138 intermediate-mass stars are identified based on their locations in the CMD, as per the fitted isochrones having masses from 1.5 to 9~M$_{\odot}$, shown in Fig.~\ref{fig:cmd}. Although some of these stars may already have reached the MS, we collectively refer to all of them as intermediate‑mass member candidates. To construct their SEDs, we used the Virtual Observatory of SED Analyser \citep[VOSA\footnote{\url{ http://svo2.cab.inta-csic.es/theory/vosa/index.php}},][]{Bayo2008A&A...492..277B}. We provided VOSA with the source coordinates, assuming extinction in the $V$ band ($A_{\rm v}$) of 1.8~mag, a distance of 2600~pc, and \textit{Spitzer}/IRAC fluxes whenever available.  VOSA then assembled fluxes from optical to infrared bands from archival databases within a 3-arcsecond matching radius. The observed fluxes are corrected for extinction using the extinction laws from \citet{Fitzpatrick1999PASP..111...63F} and \citet{Indebetouw2005ApJ...619..931I}.

The optical part of each SED was then fitted with the Kurucz stellar model \citep{Castelli1997A&A...318..841C} to obtain the parameters of the photosphere.  In the fitting, we fixed \textit{[Fe/H]} as 0.0~dex, which is the cluster metallicity, while temperature is treated as a free parameter, with surface gravity $log\,g$ varying between 3 and 5, appropriate for MS stars. The best-fitting effective temperatures of the 118 objects with reliable SEDs span $\approx 5000$--26000~K. Such a wide range is expected for the mass interval selected here, as some of these objects are still contracting and some might have already contracted onto the zero-age MS. The sample therefore covers the early F to mid B spectral type objects. A similar temperature spread is observed by \citet{Guzman2021A&A...650A.182G} for 209 bona fide sample of Herbig Ae/Be stars.

The SED in the infrared, where the emission comes mainly from thermalized dust, was used to select disk-bearing candidates by the infrared spectral index ($\alpha$), the slope of the $log$-$log$ SED in infrared wavelengths, as 
\[ \alpha = \frac{d \log \lambda F_\lambda}{ d \log \lambda}, \] 
where $\lambda$ is the wavelength and $F_\lambda$ is the observed flux density, to quantify excess emission arising from thermalized circumstellar dust. Because the amounts of dust dissipate as a young object ages, the index serves as an indicator of its evolutionary stage, from Class~0 (cloud cores) and I (deeply embedded protostars), with $\alpha > 0.3$, signifying a flat or rising SED toward long wavelengths, to Class~II (T~Tauri stars), with $0.3 > \alpha > -1.6$, then to Class~III, with $ \alpha < -1.6$ \citep[e.g.,][]{Lada1987IAUS..115....1L, Dunham2014prpl.conf..195D, Contreras2025ApJ...987...23C, Neha2025ApJS..278...10N}.

Nominally, the $\alpha$ index is computed in the wavelength range between about $2 \micron$ and $20 \micron$.  However, the majority of our sources lack \textit{WISE} $W3$ and $W4$ band data or have only upper limits. Therefore, we limited the slope fitting from the 2MASS $J$ band to the \textit{WISE} $W2$ band or Spitzer/\textit{IRAC} $I2$ band whenever available. Because our targets exhibit no sign of a flat/rising trend in SEDs, this long wavelength cutoff does not affect the derived index.  

The best-fit SED for each source is then chosen based on the minimal reduced $\chi^2$ value, calculated as:
\begin{equation} 
    \chi^2_r = \frac{1}{N-n_p}\sum_{i=0}^{N} \frac{(F_{o,i}-M_d \, F_{m,i})^2}{\sigma_{o,i}^2}, 
\end{equation}
where $N$ is the total number of photometric data points, $n_{\mathrm{p}}$ is the total number of model parameters, $F_{o,i}$ is the observed flux, $F_{m,i}$ is the model flux, $\sigma_{o,i}$ is the observational error in the flux, and $M_d$ is the dilution factor given as $(R/D)^2$ for an object of radius $R$ and at a distance of $D$. 

Of the 138 objects, except for those 20 that do not have enough data coverage or have ambiguous counterparts within the 3-arcsec matching radius, the remaining 118 objects have reliable SEDs, and hence securely derived $\alpha$ indices.  Among these, there are 24 disk-bearing candidates, including 10 Class~II ($-1.6 \leq \alpha < 0.3$) and 14 Class~III ($-2.4 \leq \alpha < -1.6$) objects.  Fig.~\ref{fig:IR_excess_SED} presents their SEDs, each being fitted by a stellar model atmosphere and the excess emission prescribed by the infrared index. Fig.~\ref{fig:CMD_RADEC} shows the CMD and spatial positions of these disk-bearing candidates. The details of disk-bearing and diskless objects are listed in Tables~\ref{tab:classII}, \ref{tab:classIII}, and \ref{tab:diskless}. 
%
 
\subsection{X-ray activity}
To explore possible additional reasons for angular momentum loss, we used the Chandra X-ray Source Catalog V2.0\footnote{\url{https://doi.org/10.25574/csc2}} \citep[CSC;][]{Evans2010ApJS..189...37E, Evans2024ApJS..274...22E} to identify X-ray point sources in Trumpler\,14.  With a 1" matching radius, there are 174 X-ray sources \citep{Salvato2018MNRAS.473.4937S} as member counterparts. Of these, 37 are in the intermediate-mass PMS regime, including 2 class~II objects, 9 class~III objects, and 26 diskless objects. Of these, 22 have $v \sin i$ measurements available. 

\begin{figure}
    \centering
    \includegraphics[width=0.9\linewidth]{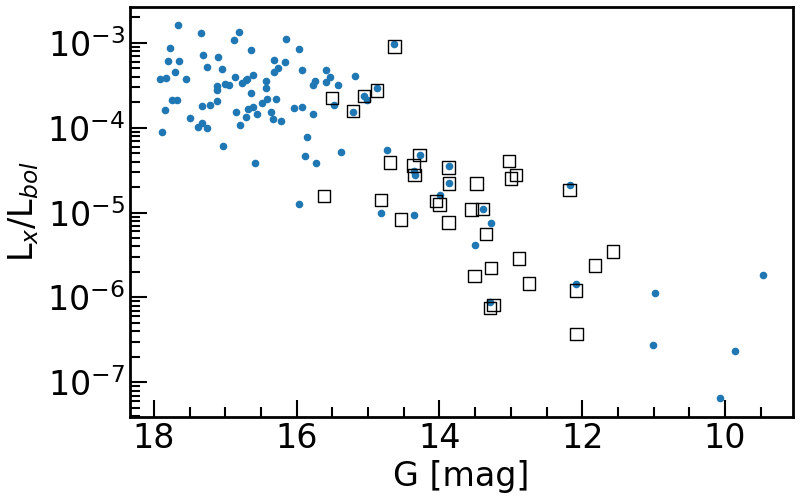}
    \caption{The ratio of X-ray to bolometric luminosity versus $G$~mag of Herbig Ae/Be stars. The blue dots represent sources with their bolometric luminosities estimated using \textit{Gaia} temperature and $\log g$, whereas the black squares represent those estimated using SED temperature and $\log g$.}
    \label{fig:xray_lum}
\end{figure}

Among the X-ray sources in our sample, the ratio of X-ray to bolometric luminosities (\(L_{\rm X}/L_{\rm bol}\)) is well correlated with the \textit{Gaia} $G$~mags, as evidenced in Fig.~\ref{fig:xray_lum}. Here the bolometric luminosities are estimated using either (1)~\textit{Gaia} temperatures and $\log g$ values, adopting a distance of 2600~pc, and $A_{\rm v} = 1.8$~mag, with the bolometric correction factor taken from the tool available on the \textit{Gaia} website\footnote{\url{https://www.cosmos.esa.int/web/gaia/dr3-bolometric-correction-tool}}, or (2)~for an additional 118 PMS objects with successful SED fitting, we also estimated the bolometric luminosities using the SED temperatures and $\log g$ values.

The ratio of X-ray to bolometric (photospheric) luminosity, $L_{\rm X}/L_{\rm bol}$, a measure of the level of magnetic activity, for our sample decreases toward brighter $G$~mag.  This is consistent with the notion that for massive OB stars, the X-rays are believed to come from colliding winds, typically with $L_{\rm X}/L_{\rm bol} \approx 10^{-7}$. In solar-type stars, with the X-rays originating from stellar dynamo magnetism, the ratio varies between $10^{-5}$ and $10^{-4}$, higher when younger and spinning faster.  In M dwarfs, with the interior being largely convective, hence magnetic, the ratio is close to the saturation value, with $L_{\rm X}/L_{\rm bol}$ reaching up to $10^{-3}$.  For T~Tauri stars, there is a known positive but scattered correlation between $L_{\rm X}$ and $L_{\rm bol}$ \citep{preibisch05b}. The T Tauri population descending along the convective Hayashi tracks has $L_{\rm X}/L_{\rm bol} > 10^{-5}$, higher than that of the present-day Sun ($L_{\rm X}/L_{\rm bol} > 10^{-6.5}$; averaging the quiescent and active Sun).  The more evolved PMS weak-lined T Tauri stars  (Class~III) are generally more magnetically active, likely as a consequence of spinning up upon contracting toward the MS.  Unlike a single MS star, for which the magnetic field loops are contained within the stellar surface, a PMS object may connect the field to the disk, rendering a complex star-disk field geometry and a larger emission volume. 
\begin{figure*}
\begin{center}
    \includegraphics[width=0.45\textwidth]{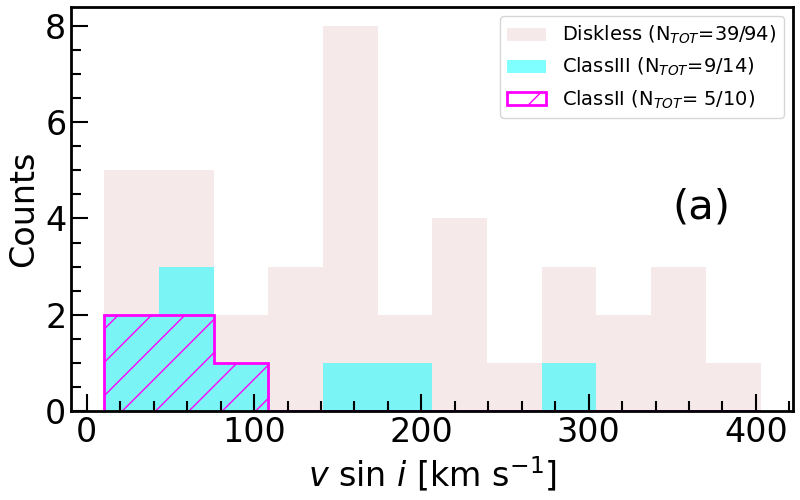}
    \includegraphics[width=0.45\textwidth]{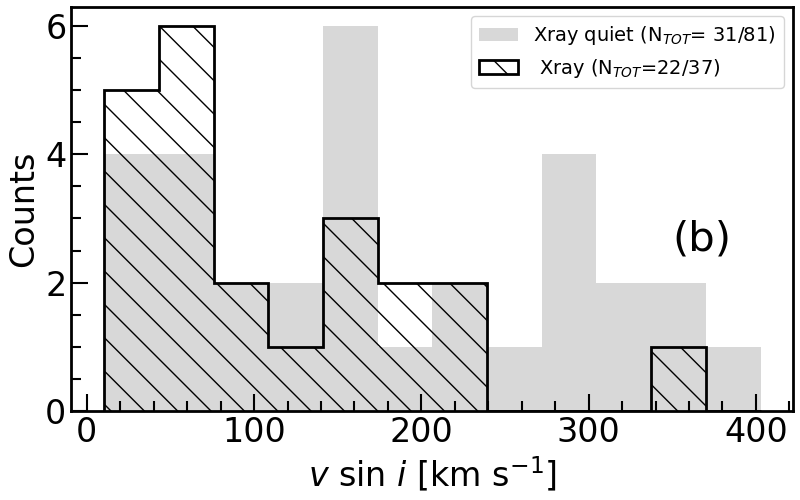}
    \caption{The $v \sin i $ distribution of (a)~disk-bearing versus diskless PMS members and (b)~X-ray versus non-X-ray PMS members.}
    \label{fig:vsini_comp}
\end{center}
\end{figure*}
%
\subsection{Binaries}
We further cross-matched the cluster members with the \textit{Gaia} variability catalog \citep{Eyer2023A&A...674A..13E} and the literature to identify photometric and spectroscopic variables or binaries. This yielded 10 eclipsing or spectroscopic binaries and 20 RS~Canum~Venaticorum (RS~CVn) variables. Of these, three binaries and with five RS~CVn variables match our intermediate-mass PMS objects. None of the binaries have X-ray detections.  All RS~CVn stars are X-ray sources, as expected, since they are chromospherically active pairs. One binary and two RS~CVn stars show evidence of dusty disks.  

One RS~CVn star has a measured $v \sin i$, and it is a slow rotator.  Of the remaining binaries, one is a slow rotator, and the other is a fast rotator. Of the three diskless RS~CVn systems, one with $v\sin i$ available is a moderate rotator. Given the extremely small samples and the possibility that $v \sin i$ may not necessarily measure rotation but could be blended with binary orbital motion, we could not draw any conclusion regarding a correlation with binarity. Therefore, we refrain from discussing this further.
%
%
\section{Discussion} \label{sec:discussion}
As discussed, a PMS object spins up when evolving toward the MS, competing with slowdown mechanisms such as disk-locking and magnetic braking.  The existence of a dusty envelope, for instance, a circumstellar disk, is diagnosed by infrared excess. However, a magnetic field is usually not readily measurable; therefore, X-ray emission is used as a proxy.  In the following, we discuss the correlation of the presence of disks or X rays with their stellar rotation in Trumpler\,14. 
\subsection{Rotation vs Disk}
Roughly 20\% (24/118) of the selected intermediate-mass PMS objects are identified as disk-bearing. Fig.~\ref{fig:CMD_RADEC} shows their location on the CMD and spatial distribution. They are predominantly located outside the central region of the cluster, as evinced in Fig.~\ref{fig:CMD_RADEC}b. The paucity of disk-bearing members in the center could be related to the presence of massive O-type members at the center. Trumpler\,14 hosts 10 known O-type stars, half of which are situated near the cluster center.  Their strong winds and UV radiation photoevaporate or photoionize nearby dust and gas, thereby inhibiting accretion and accelerating disk dispersal in neighboring PMS objects. The UV luminosity of Trumpler\,14 is estimated to be $\approx 2 \times 10^{50}$ photons~s$^{-1}$ \citep{Smith2008hsf2.book..138S}, sufficient to evaporate a 50~M$_{\rm Jup}$ disk within $10^5$~yr. \citet{Mesa2016ApJ...825L..16M} reported the general absence of submillimeter emission in the region, suggesting clearing of circumstellar dust by massive members.  Our finding of the paucity of disk-bearing Herbig Ae/Be stars near O-type stars is consistent with this scenario. Additionally, \citet{Polnitzky2026A&A...707A.216P} reported that the average disk dispersal time of stars from the hydrogen-burning limit to 8~M$\odot$ across 33 young clusters is $5.8 \pm 0.3$~Myr.  Trumpler\,14 has an age of 1--7.9~Myrs, so as to have already lost the disks in many members.  

Of the 118 intermediate-mass PMS objects, 53 have available $v \sin i $ measurements, with 23 of them having $v \sin i < 100$~km~s$^{-1}$. The $v \sin i$ distributions of disk-bearing Class~II and Class~III and diskless objects from different studies are compared in Fig.~\ref{fig:vsini_comp}a. All 5/10 Class~II objects with $v \sin i $ data are slow rotators. Among 9/14 Class~III disk-bearing objects with $v \sin i $, all but three (star\#36, star\#59, and star\#94) are slow rotators. Star~59 has two $v \sin i $ measurements reported in the literature, 383.2~km~s$^{-1}$ \citep{Randich2022A&A...666A.121R} and $165 \pm 24.75$~km~s$^{-1}$ \citep{Hourihane2023A&A...676A.129H}. It is unclear whether the two values differ due to binarity, but in any case, the star is a relatively fast rotator. The details of $v \sin i$ measurements of the intermediate-mass members are summarized in Appendix~\ref{sec:vsini}. 

A two-sample Anderson–Darling test \citep[AD statistics = 1.0, p-value = 0.126;][]{Anderson01121954} does not indicate a statistically significant difference between the $ v \sin i$ distributions of disk-bearing objects (5 Class~II and 9 Class~III) versus diskless objects (39). However, the small sample size of disk-bearing objects, along with three Class~III objects having $ v \sin i$ measurements that fall in the tail of the distribution, reduces the statistical power of the test. As shown in Fig~\ref{fig:vsini_comp}a, the $ v \sin i$ distribution of disk-bearing objects peaks below 100~km~$s^{-1}$ and most disk-bearing objects are slow rotators, which is in agreement with \citet{Bastian2020MNRAS.495.1978B} and \citet{Bu2025ApJ...979...29B}. 

The Class~II and Class~III objects also span similar ranges in magnitude and age, as judged from their overlapping distributions relative to the two fitted isochrones in Fig.~\ref{fig:CMD_RADEC}a, and show no systematic difference in their spatial distributions in Fig.~\ref{fig:CMD_RADEC}b. We therefore can consider them as drawn from the same coeval population, with similar ranges in magnitude and spatial distribution. None of the Class~II stars are fast rotators, whereas only three Class~III objects exhibit rapid rotation. This supports the notion that stars retaining their disks for longer periods spin more slowly than those that have already dispersed their disks.
%
\subsection{Rotation vs X-ray}
The rotation of X-ray versus non-X-ray sources is compared in Fig.~\ref{fig:vsini_comp}b. Even though the sample sizes are small, it appears that the X-ray sources tend to have relatively smaller $v \sin i$ values, except for one outlier beyond 300~km~s$^{-1}$. A two-sample Anderson–Darling test \citep{Anderson01121954} indicates that the $v \sin i$ of the two samples differ at the 95\% confidence level ($AD = 2.092$, $p = 0.045$). The slower rotation of X-ray sources suggests a possible connection between X-ray activity and angular momentum loss in intermediate-mass PMS objects. If the X rays in these objects originate from magnetic activities associated with residual magnetic fields, magnetic star–disk interaction or magnetically driven winds could contribute to angular momentum removal, resulting in slower rotation. However, given the uncertain origin of X rays in Herbig Ae/Be stars, this interpretation remains tentative.

For Herbig Ae/Be stars, a convective track exists, though not as prominent as for T Tauri stars, particularly when young and toward the low-mass end, suggesting that a magnetic field could be responsible for the X rays.  Several Herbig Ae/Be stars are indeed known X-ray sources \citep{damiani94,hamaguchi05}, but the origin of their X-ray emission, if detected, is an unsettled issue. It could be attributed to wind-driven shocks as in OB stars \citep{Zinnecker1994A&A...292..152Z}, to magnetic dynamos as in low-mass stars \citep{Hamaguchi2005ApJ...618..360H, Skinner2004ApJ...614..221S, Stelzer2006A&A...457..223S}, or to an active X-ray companion \citep{Zinnecker1994A&A...292..152Z, Skinner2004ApJ...614..221S, Stelzer2006A&A...457..223S, Stelzer2009A&A...499..529S}.  

\citet{hamaguchi05} demonstrated that the X-ray luminosities of Herbig Ae/Be stars are higher than those of T Tauri stars, thereby ruling out unresolved low-mass companions as the X-ray source. \citet{Hubrig2009A&A...502..283H} reported a correlation between magnetic field strengths and X-ray luminosities of a sample of Herbig Ae/Be stars, which decay with time. They also identified several X-ray active Herbig Ae/Be stars in binary systems, with the X-ray emissions attributed to the Herbig Ae/Be stars themselves. These results suggest that some Herbig Ae/Be stars may possess magnetic fields to generate X rays, perhaps through radiative shear \citep{Tout1995MNRAS.272..528T}. Similarly, diskless PMS objects with residual magnetic fields may continue to produce X rays. Such X-ray activities could potentially lead to angular momentum loss and may explain the smaller $v \sin i$ values observed among X-ray sources. Furthermore, \citet{Alecian2013MNRAS.429.1027A} investigated 70 Herbig Ae/Be stars located in different regions of the galaxy, and found five magnetic Herbig Ae/Be stars that are slower rotators compared to their non-magnetic counterparts, indicating efficient early braking and subsequent angular momentum loss. Such a scenario has been proposed in a particular PMS binary, in which the 6~M$_\odot$ companion was found to be magnetic and slowly rotating, in contrast to the 12~M$_{\odot}$ non-magnetic primary  \citep{Shultz2021MNRAS.504.3203S}. This finding also supports the magnetic braking mechanism, leading to slower rotation of the magnetic companion. 
Thus, these results suggest that magnetic fields and winds may provide possible pathways for angular momentum loss in intermediate-mass PMS objects.
%
%
\section{Summary \& Conclusion  }\label{sec:summary}
In this study, we investigated possible causes for early loss of angular momentum in intermediate-mass PMS members in the young cluster Trumpler\,14 using X-ray, optical, and IR data.  

\begin{enumerate}
    \item We identified 342 cluster members using the ML-MOC algorithm within 5 arcmin of the cluster center.
    \item Among 118 intermediate-mass PMS members, 24 (10 Class~II and 14 Class~III) are found by their infrared excess with substantial dusty disks, the majority of which are preferentially located away from the most massive member stars, signifying disk clearing in the massive stars' vicinities. 
    \item Of the 24 disk-bearing Herbig Ae/Be candidates, 14 have available $v \sin i$ measurements. Class~II objects are preferentially slow rotators, and the majority of Class~III objects are also slow rotators. 
    \item Of the 118 intermediate-mass PMS objects, 37 are X-ray sources, including eight disk-bearing objects, which by and large rotate slowly. 
\end{enumerate}

Massive and luminous members of Trumpler\,14 play a destructive role in the dusty envelopes around adjacent Herbig Ae/Be stars, with about a fraction of 20\% remaining disk-bearing.  These disk objects tend to be slow rotators, manifesting the consequence of star-disk magnetic interactions.  Our analysis suggests that X-ray emission, as a proxy for young stellar magnetism, can also lead to angular-momentum loss.  Taken together, our analysis argues that a combination of magnetic activity and disk-related braking contributes to slowing down intermediate-mass PMS objects.
The remaining slow rotators, some of which may well spin fast but are disguised by inclination and lack disk or X‑ray signatures, require additional explanations. Therefore, rotational velocities of the intermediate-mass population of a cluster during their MS phase depend on a multitude of early angular momentum evolution.
Larger spectroscopic samples and direct magnetic diagnostics of intermediate-mass PMS objects will be required to quantify the relative roles of these mechanisms.
%
%
\begin{acknowledgments}
    We thank the anonymous referee for carefully reading the manuscript and for constructive comments. 
    KKR and WPC acknowledge funding from the National Science and Technology Council of Taiwan (NSTC~115-2112-M-008-013). This work includes data from the ESO Science Archive Facility \citep{https://doi.org/10.18727/archive/25} and the third data release from the European Space Agency (ESA) mission {\it Gaia} \citep[\url{https://www.cosmos.esa.int/gaia};][]{Gaia2023A&A...674A...1G}, processed by the {\it Gaia} Data Processing and Analysis Consortium (DPAC, \url{https://www.cosmos.esa.int/web/gaia/dpac/consortium}). We also used \textit{Chandra} X-ray data \citep{csc2}, and the Astrophysics Data System (ADS) governed by NASA (\url{https://ui.adsabs.harvard.edu}).
    The following tools are also used in this work: 
    \textsc{Astropy} \citep{2018AJ....156..123A}; 
    \textsc{Astroquery} \citep{Ginsburg2019AJ....157...98G};
    \textsc{Matplotlib} \citep{Hunter:2007};
    \textsc{NumPy} \citep{2020Natur.585..357Harris}; 
    \textsc{SciPy} \citep{2020SciPy-NMeth}; 
    \textsc{topcat} \citep{2005ASPC..347...29TOPCAT}. 
\end{acknowledgments}
%
%
\appendix
\section{Details of \texorpdfstring{$v \sin i$}{v sin i} measurements} \label{sec:vsini} 
\begin{figure}
    \centering
    \includegraphics[width=0.45\linewidth]{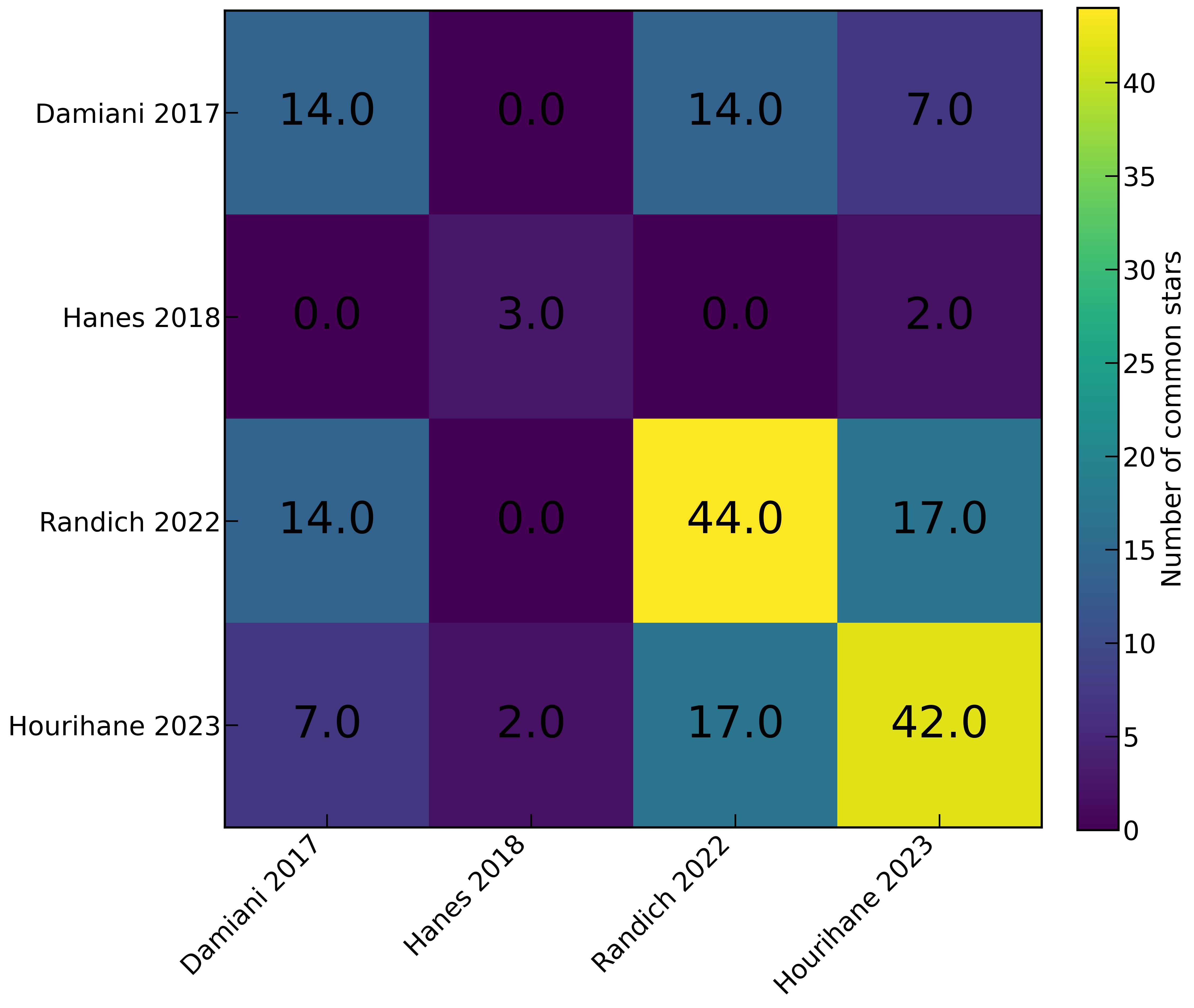}
    \includegraphics[width=0.45\linewidth]{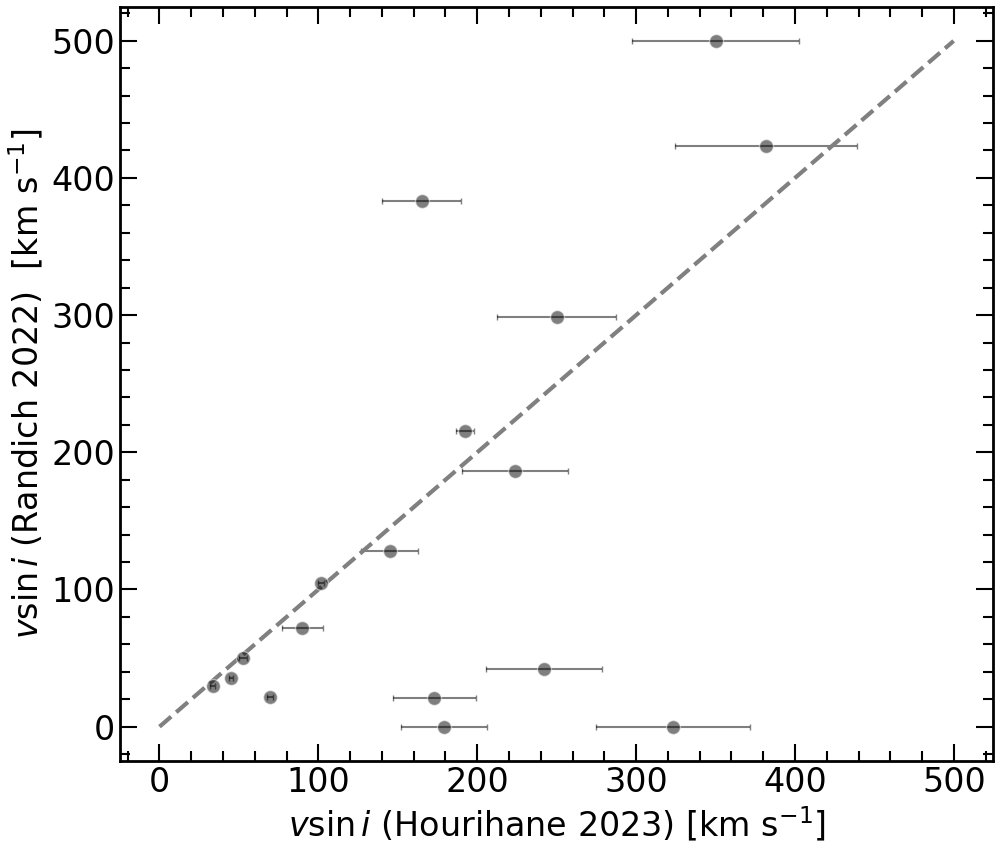}
    \caption{Left panel: a matrix representation of $v \sin i$ measurements of the MSTO members from four different catalogs. Right panel: Crossmatch between $v \sin i$ measurements from \citet{Hourihane2023A&A...676A.129H} and \citet{Randich2022A&A...666A.121R} for 17 common sources. The grey dashed line shows a one-to-one correlation between $v \sin i$ measurements from \citet{Hourihane2023A&A...676A.129H} and \citet{Randich2022A&A...666A.121R}.}
    \label{fig:vsini_xmatch}
\end{figure}

When cross-matching the 138 MS turn-on members with SIMBAD, 70 were found to have $v \sin i$ measurements. A matrix representation of the majority of the $v \sin i$ values from four different catalogs is shown in Fig.~\ref{fig:vsini_xmatch}. The majority of them are observed in \citet{Randich2022A&A...666A.121R} and \citet{Hourihane2023A&A...676A.129H}. As we can see, the maximum number of common sources is between the works by \citet{Hourihane2023A&A...676A.129H} and \citet{Randich2022A&A...666A.121R}, with the measurements consistent with each other within the errors. In our sample, if a source has multiple measurements, we adopted the median value. Tables~\ref{tab:classII}, ~\ref{tab:classIII}, ~\ref{tab:diskless} show the median $v \sin i$ value and the total number of $v \sin i$ measurements available in the literature.  

\begin{figure}
\begin{center}
    \includegraphics[width=0.95\linewidth]{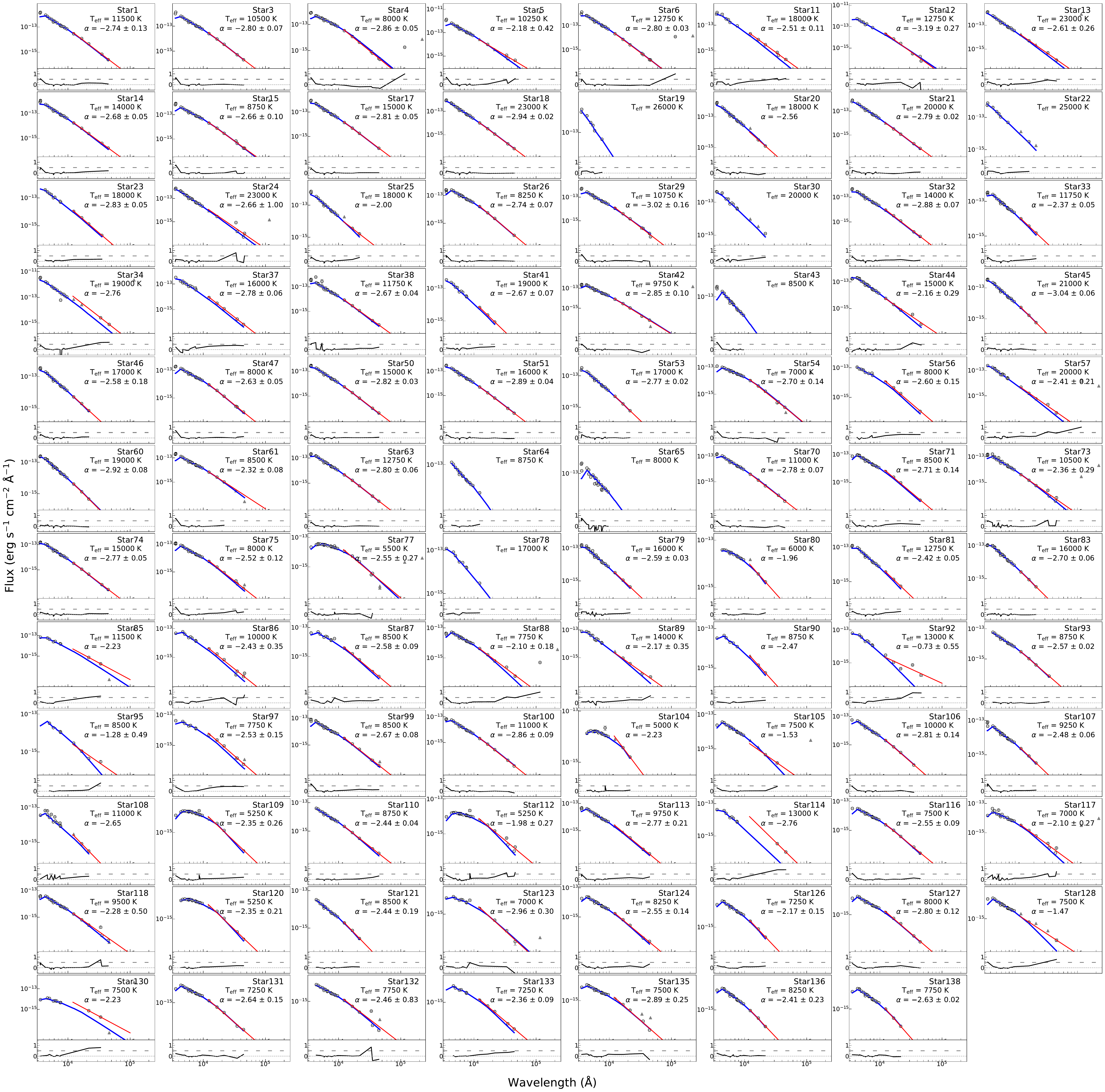}
    \caption{Same as Figure \ref{fig:IR_excess_SED} but for objects without significant IR excess.}
    \label{fig:without_excess_SEDs}
\end{center}
\end{figure}

\centerwidetable
\begin{deluxetable*}{llllllllll}
\tablecaption{Details of the Class II intermediate-mass disk-bearing objects.}
\label{tab:classII}
\tablewidth{0.pt}
\tabletypesize{\tiny}
\tablehead{
\colhead{Name} &
\colhead{Gaia ID} &
\colhead{RA} &
\colhead{DEC} &
\colhead{$G$} &
\colhead{$\alpha$} &
\colhead{Chandra ID} &
\colhead{Chandra Flux} &
\colhead{$N_{v\sin i}$} &
\colhead{$v\sin i$} \\
&
&
\colhead{(deg)} &
\colhead{(deg)} &
\colhead{(mag)} &
&
&
\colhead{(erg s$^{-1}$ cm$^{-2}$)} &
&
\colhead{(km s$^{-1}$)}
}
\startdata
        Star2 & 5350363940295884416 & 160.999381 & $-$59.529781 & 11.6 & $-$0.73 $\pm$ 0.07 & ~ & ~ & ~ & ~ \\ 
        Star16 & 5350363459259486208 & 160.943335 & $-$59.578445 & 12.19 & $-$0.94 $\pm$ 0.05 & ~ & ~ & 2.0 & 58.80 $\pm$ 10.00 \\ 
        Star31 & 5350387481038542080 & 160.974694 & $-$59.511001 & 12.87 & $-$0.82 $\pm$ 0.06 & ~ & ~ & ~ & ~ \\ 
        Star52 & 5350363841537227008 & 160.958878 & $-$59.568582 & 13.3 & $-$0.81 $\pm$ 0.04 & ~ & ~ & 1.0 & 82.90     \\ 
        Star58 & 5350362978223239168 & 161.041437 & $-$59.576725 & 13.39 & $-$1.60 $\pm$ 0.15 & 2CXO J104409.9-593436 & 4.121e$-$15 & 3.0 & 32.96     \\ 
        Star66 & 5350363051263290752 & 161.002488 & $-$59.567908 & 13.63 & $-$0.35 $\pm$ 0.10 & ~ & ~ & ~ &  ~  \\ 
        Star72 & 5350363154342551680 & 161.033459 & $-$59.562208 & 13.79 & $-$0.61 $\pm$ 0.11 & ~ & ~ & ~ & ~ \\ 
        Star91 & 5350362432793015040 & 161.008642 & $-$59.631111 & 14.28 & $-$0.99 $\pm$ 0.07 & 2CXO J104402.0-593751 & 7.894e$-$15 & 3.0 & 69.30 $\pm$ 1.96 \\ 
        Star101 & 5350364219494459264 & 160.947109 & $-$59.518828 & 14.52 & $-$0.59 $\pm$ 0.10 & ~ & ~ & ~ &  ~  \\ 
        Star119 & 5350387687196971776 & 160.940784 & $-$59.497154 & 14.84 & $-$1.45 $\pm$ 0.10 & ~ & ~ & 3.0 & 35.20 $\pm$ 1.39 \\ 
\enddata
\end{deluxetable*}

\begin{deluxetable*}{llllllllll}
\tablecaption{Details of the Class III intermediate-mass disk-bearing objects.}
\label{tab:classIII}
\tablewidth{0pt}
\tabletypesize{\tiny}
\tablehead{
\colhead{Name} &
\colhead{Gaia ID} &
\colhead{RA} &
\colhead{DEC} &
\colhead{$G$} &
\colhead{$\alpha$} &
\colhead{Chandra ID} &
\colhead{Chandra Flux} &
\colhead{$N_{v\sin i}$} &
\colhead{$v\sin i$} \\
&
&
\colhead{(deg)} &
\colhead{(deg)} &
\colhead{(mag)} &
&
&
\colhead{(erg s$^{-1}$ cm$^{-2}$)} &
&
\colhead{(km s$^{-1}$)}
}
\startdata
        Star36 & 5350362570226891648 & 161.028383 & $-$59.603185 & 13.0 & $-$2.25 $\pm$ 0.06 & 2CXO J104406.8-593611 & 1.682e$-$14 & 2.0 & 174.69 $\pm$ 46.80 \\ 
        Star39 & 5350363665417895168 & 160.922605 & $-$59.564684 & 13.03 & $-$2.19 $\pm$ 0.15 & 2CXO J104341.4-593352 & 2.265e$-$14 & 2.0 & 55.55 $\pm$ 9.60 \\ 
        Star55 & 5350363532299502208 & 160.902535 & $-$59.580841 & 13.34 & $-$2.32 $\pm$ 0.21 & ~ & ~ & 1.0 & 42.10     \\ 
        Star59 & 5350363119982842240 & 161.053417 & $-$59.55897 & 13.43 & $-$1.61 $\pm$ 0.23 & ~ & ~ & 2.0 & 274.10 $\pm$ 24.75 \\ 
        Star62 & 5350386690764583424 & 161.037298 & $-$59.525654 & 13.48 & $-$1.96 $\pm$ 0.07 & 2CXO J104408.9-593132 & 7.996e$-$15 & 2.0 & 38.35 $\pm$ 15.70 \\ 
        Star76 & 5350363807177532416 & 160.979602 & $-$59.559551 & 13.87 & $-$2.01 $\pm$ 0.06 & 2CXO J104355.1-593334 & 5.268e$-$15 & 1.0 & 51.70  \\ 
        Star84 & 5350386450246441088 & 161.135591 & $-$59.550967 & 14.05 & $-$2.06 $\pm$ 0.07 & 2CXO J104432.5-593303 & 9.521e$-$15 & 2.0 & 49.10 $\pm$ 10.00 \\ 
        Star94 & 5350383392217747456 & 161.141297 & $-$59.565293 & 14.35 & $-$1.96 $\pm$ 0.13 & 2CXO J104433.9-593355 & 4.326e$-$15 & 1.0 & 172.70  \\ 
        Star103 & 5350363669738519936 & 160.918193 & $-$59.556791 & 14.54 & $-$1.79 $\pm$ 0.02 & 2CXO J104340.3-593324 & 1.063e$-$15 & ~ &  ~  \\ 
        Star122 & 5350387412319080960 & 161.006986 & $-$59.517116 & 15.06 & $-$2.05 $\pm$ 0.10 & 2CXO J104401.6-593101 & 1.916e$-$14 & 3.0 & 104.60 $\pm$ 9.05 \\ 
        Star125 & 5350362501507335168 & 160.99099 & $-$59.626082 & 15.18 & $-$1.70 $\pm$ 0.16 & ~ & ~ & ~ &  ~  \\ 
        Star129 & 5350363326141059584 & 160.918822 & $-$59.594407 & 15.51 & $-$1.63 $\pm$ 0.35 & 2CXO J104340.6-593539 & 1.225e$-$14 & ~ &  ~  \\ 
        Star134 & 5350363326141055744 & 160.918572 & $-$59.595863 & 15.64 & $-$1.80 $\pm$ 0.12 & ~ & ~ & ~ &  ~  \\ 
        Star137 & 5350363326141072512 & 160.92921 & $-$59.59194 & 15.75 & $-$2.34 $\pm$ 0.11 & ~ & ~ &  & \\ 
\enddata
\end{deluxetable*}

\begin{table*}[!ht]
    \centering
    \caption{Details of the intermediate-mass diskless objects.}
    \label{tab:diskless}
    \tiny
    \begin{tabular}{lllllllll}
    \hline
    \\
        Name & Gaia ID & RA & DEC & G & Chandra ID & Chandra Flux & N$_{v \sin i}$ & $v \sin i$ \\ 
         &  & (deg) & (deg) & (mag) &   &  (erg~s$^{-1}$~cm$^{-2}$) &  & (km~s$^{-1}$)\\ 
         \\
        \hline
        \\ 
        Star1 & 5350363905936139904 & 160.99348 & $-$59.550449 & 11.57  & 2CXO J104358.4-593301 & 1.3e-14 & 2.0 & 237.50 $\pm$ 36.64 \\ 
        Star3 & 5350363841537240064 & 160.962005 & $-$59.564063 & 11.8 & ~ & ~ & 1.0 & 211.74 $\pm$ 31.76 \\ 
        Star4 & 5350387824635992832 & 160.974477 & $-$59.473143 & 11.82  & 2CXO J104353.8-592823 & 4.123e$-$15 & ~ &  ~  \\ 
        Star5 & 5350363768497176320 & 160.983389 & $-$59.578048 & 11.82   & ~ & ~ & 1.0 & 114.28 $\pm$ 17.14 \\ 
        Star6 & 5350386347167198976 & 161.134227 & $-$59.566514 & 11.83   & ~ & ~ & 1.0 & 305.00 $\pm$ 45.75 \\ 
        Star11 & 5350363631058143488 & 160.910959 & $-$59.579001 & 12.02   & ~ & ~ & 2.0 & 402.75 $\pm$ 57.30 \\ 
        Star12 & 5350362497181876608 & 160.967543 & $-$59.616269 & 12.08   & 2CXO J104352.1-593658 &   ~& 1.0 & 106.40  \\ 
        Star13 & 5350363875897031296 & 160.96685 & $-$59.544508 & 12.08   & 2CXO J104352.0-593239 & 4.094e$-$15 & 1.0 & 156.90 $\pm$ 23.54 \\ 
        Star14 & 5350363841537241088 & 160.953286 & $-$59.559777 & 12.09   & 2CXO J104348.7-593335 & 4.309e$-$15 & 1.0 & 345.00 $\pm$ 10.00 \\ 
        Star15 & 5350363459259473664 & 160.931357 & $-$59.567637 & 12.18   & 2CXO J104343.5-593403 & 2.362e$-$14 & 2.0 & 48.24 $\pm$ 31.48 \\ 
        Star17 & 5350364082055448192 & 160.923228 & $-$59.531706 & 12.24   & ~ & ~ & ~ &  ~  \\ 
        Star18 & 5350362810745171072 & 161.101942 & $-$59.590341 & 12.36   & ~ & ~ & ~ &  ~  \\ 
        Star19 & 5350363905936127872 & 160.97993 & $-$59.544319 & 12.39   & ~ & ~ & ~ &  ~  \\ 
        Star20 & 5350363905936142080 & 160.995344 & $-$59.546539 & 12.4   & ~ & ~ & ~ &  ~  \\ 
        Star21 & 5350363944616554624 & 160.987955 & $-$59.535568 & 12.43   & ~ & ~ & ~ &  ~  \\ 
        Star22 & 5350363905936138112 & 160.991624 & $-$59.541887 & 12.49   & ~ & ~ & ~ &  ~  \\ 
        Star23 & 5350362432787875072 & 161.013267 & $-$59.627163 & 12.53   & ~ & ~ & 1.0 & 355.60     \\ 
        Star24 & 5350386617718203392 & 161.074648 & $-$59.545545 & 12.57   & ~ & ~ & ~ &  ~  \\ 
        Star25 & 5350363905936133248 & 160.986328 & $-$59.544586 & 12.65   & ~ & ~ & ~ &  ~  \\ 
        Star26 & 5350363910241982336 & 160.984159 & $-$59.541068 & 12.67   & ~ & ~ & 1.0 & 40.00 $\pm$ 6.00 \\ 
        Star29 & 5350362600261090816 & 160.979147 & $-$59.60669 & 12.75   & 2CXO J104354.9-593624 & 1.61e$-$15 & 2.0 & 181.00 $\pm$ 31.13 \\ 
        Star30 & 5350363115657162752 & 161.050612 & $-$59.564191 & 12.81   & ~ & ~ & 2.0 & 141.85 $\pm$ 36.30 \\ 
        Star32 & 5350386312807445248 & 161.111448 & $-$59.564138 & 12.88   & ~ & ~ & 1.0 & 150.00 $\pm$ 22.50 \\ 
        Star33 & 5350386347167213312 & 161.13538 & $-$59.558109 & 12.89   & 2CXO J104432.4-593329 & 3.346e$-$15 & 1.0 & 50.00 $\pm$ 7.50 \\ 
        Star37 & 5350363463580099968 & 160.949828 & $-$59.56642 & 13.02   & ~ & ~ & 2.0 & 205.00 $\pm$ 33.56 \\ 
        Star38 & 5350363463564711168 & 160.940716 & $-$59.576653 & 13.02   & ~ & ~ & 1.0 & 337.00 $\pm$ 50.55 \\ 
        Star41 & 5350386347167197824 & 161.122663 & $-$59.5622 & 13.1   & ~ & ~ & 2.0 & 161.50 $\pm$ 48.45 \\ 
        Star42 & 5350364077734764672 & 160.932189 & $-$59.537891 & 13.14   & ~ & ~ & 1.0 & 250.00 $\pm$ 37.50 \\ 
        Star43 & 5350363463580074880 & 160.936575 & $-$59.571032 & 13.17   & ~ & ~ & 1.0 & 293.00 $\pm$ 43.95 \\ 
        Star44 & 5350386725124362496 & 161.08101 & $-$59.519793 & 13.18   & ~ & ~ & 1.0 & 150.00 $\pm$ 22.50 \\ 
        Star45 & 5350386450246447744 & 161.11857 & $-$59.539739 & 13.19   & ~ & ~ & 1.0 & 168.00 $\pm$ 25.20 \\ 
        Star46 & 5350363974655589120 & 160.967819 & $-$59.543355 & 13.24   & 2CXO J104352.2-593236 & 1.582e$-$15 & 1.0 & 129.00 $\pm$ 19.35 \\ 
        Star47 & 5350363738457983232 & 160.899771 & $-$59.554983 & 13.26   & ~ & ~ & 1.0 & 44.00 $\pm$ 6.60 \\ 
        Star50 & 5350363188702311936 & 161.017901 & $-$59.546633 & 13.28   & 2CXO J104404.2-593247 & 3.1e$-$15 & 1.0 & 209.00 $\pm$ 31.35 \\ 
        Star51 & 5350387206160734336 & 161.062327 & $-$59.488333 & 13.29   & 2CXO J104414.9-592917 & 1.215e$-$15 & 1.0 & 150.00 $\pm$ 22.50 \\ 
        Star53 & 5350363944616573824 & 161.002497 & $-$59.531247 & 13.31   & ~ & ~ & 1.0 & 207.00 $\pm$ 31.05 \\ 
        Star54 & 5350363463580073472 & 160.954588 & $-$59.579699 & 13.31   & ~ & ~ & ~ &  ~  \\ 
        Star56 & 5350364047695662592 & 160.918316 & $-$59.549572 & 13.35   & 2CXO J104340.3-593258 & 2.343e$-$15 & 1.0 & 39.60     \\ 
        Star57 & 5350386759484114560 & 161.085281 & $-$59.513096 & 13.36   & ~ & ~ & 2.0 & 274.30 $\pm$ 37.50 \\ 
        Star60 & 5350386587685344640 & 161.064249 & $-$59.548176 & 13.44   & ~ & ~ & 1.0 & 300.00 $\pm$ 45.00 \\ 
        Star61 & 5350363429220300160 & 160.934548 & $-$59.588486 & 13.45   & ~ & ~ & ~ &  ~  \\ 
        Star63 & 5350364082055463424 & 160.93523 & $-$59.530538 & 13.51   & 2CXO J104344.4-593149 & 1.39e$-$15 & ~ &  ~  \\ 
        Star64 & 5350363772817768448 & 160.967325 & $-$59.563495 & 13.55   & 2CXO J104352.1-593348 & 3.962e$-$15 & ~ &  ~  \\ 
        Star65 & 5350363463580073728 & 160.93507 & $-$59.571333 & 13.56   & ~ & ~ & ~ &  ~  \\ 
        Star70 & 5350363051263307648 & 161.007072 & $-$59.562513 & 13.74   & ~ & ~ & ~ &  ~  \\ 
        Star71 & 5350362668980567808 & 161.079196 & $-$59.609239 & 13.78   & ~ & ~ & ~ &  ~  \\ 
        Star73 & 5350387717230369536 & 160.963728 & $-$59.483283 & 13.79   & ~ & ~ & 1.0 & 94.00 $\pm$ 14.10 \\ 
        Star74 & 5350362810745165056 & 161.09292 & $-$59.587909 & 13.82   & ~ & ~ & ~ &  ~  \\ 
        Star75 & 5350363360501081728 & 160.96555 & $-$59.595846 & 13.87   & 2CXO J104351.7-593545 & 5.405e$-$15 & ~ &  ~  \\ 
        Star77 & 5350387481038536320 & 160.982316 & $-$59.518483 & 13.87   & 2CXO J104355.7-593106 & 8.446e$-$15 & ~ &  ~  \\ 
        Star78 & 5350363807177553664 & 160.981258 & $-$59.55095 & 13.91   & ~ & ~ & ~ &  ~  \\ 
        Star79 & 5350386381526917760 & 161.077855 & $-$59.551817 & 13.94   & ~ & ~ & ~ &  ~  \\ 
        Star80 & 5350363905936141568 & 160.995233 & $-$59.551953 & 13.98   & ~ & ~ & ~ &  ~  \\ 
        Star81 & 5350363188702408448 & 161.037036 & $-$59.55564 & 14.0   & 2CXO J104408.9-593320 & 6.218e$-$15 & ~ &  ~  \\ 
        Star83 & 5350386415886683392 & 161.08279 & $-$59.541008 & 14.02   & ~ & ~ & ~ &  ~  \\ 
        Star85 & 5350363429220314880 & 160.927878 & $-$59.578788 & 14.08   & ~ & ~ & 1.0 &   ~    \\ 
        Star86 & 5350362810745141376 & 161.095216 & $-$59.598465 & 14.14   & ~ & ~ &  & 59.10     \\ 
        Star87 & 5350386622045115264 & 161.079412 & $-$59.538951 & 14.23   & ~ & ~ & 2.0 & 136.35 $\pm$ 18.00 \\ 
        Star88 & 5350387584117846784 & 161.03695 & $-$59.484927 & 14.24   & ~ & ~ & ~ &  ~  \\ 
        \\
        \hline
    \end{tabular}
\end{table*}

\begin{table*}[!ht]
    \centering
    \caption{Continuation of Table~\ref{tab:diskless} of diskless objects.}
    \tiny
    \begin{tabular}{lllllllll}
    \hline
    \\
        Name & Gaia ID & RA & DEC & G & Chandra ID & Chandra Flux & N$_{v \sin i}$ & $v \sin i$ \\ 
         &  & (deg) & (deg) & (mag) &   &  (erg~s$^{-1}$~cm$^{-2}$) &  & (km~s$^{-1}$)\\ 
         \\
        \hline
        \\ 
        Star89 & 5350383392229689472 & 161.142108 & $-$59.574979 & 14.24   & ~ & ~ & 1.0 & 56.50     \\ 
        Star90 & 5350364219494466944 & 160.953572 & $-$59.518565 & 14.27   & ~ & ~ &  &   ~    \\ 
        Star92 & 5350363532299499264 & 160.889602 & $-$59.577309 & 14.3   & ~ & ~ & 1.0 & 343.60     \\ 
        Star93 & 5350363051263299328 & 161.020358 & $-$59.571615 & 14.3   & ~ & ~ & ~ &  ~  \\ 
        Star95 & 5350364150774922752 & 160.914391 & $-$59.529459 & 14.36   & ~ & ~ &  &   ~    \\ 
        Star97 & 5350363669738511488 & 160.903103 & $-$59.554228 & 14.36   & 2CXO J104336.7-593315 & 5.891e$-$15 & 2.0 & 41.30 $\pm$ 12.70 \\ 
        Star99 & 5350363360500832256 & 160.949265 & $-$59.591056 & 14.41   & 2CXO J104347.8-593528 &   ~& ~ &  ~  \\ 
        Star100 & 5350364219494467200 & 160.938318 & $-$59.51195 & 14.47   & ~ & ~ & ~ &  ~  \\ 
        Star104 & 5350363875897005568 & 160.945893 & $-$59.545086 & 14.57   & ~ & ~ & ~ &  ~  \\ 
        Star105 & 5350387717230358528 & 160.953702 & $-$59.491889 & 14.58   & ~ & ~ & ~ &  ~  \\ 
        Star106 & 5350363154342665728 & 161.022362 & $-$59.56518 & 14.58   & ~ & ~ & ~ &  ~  \\ 
        Star107 & 5350363463580071040 & 160.944329 & $-$59.576871 & 14.61   & ~ & ~ & ~ &  ~  \\ 
        Star108 & 5350386656404822912 & 161.039523 & $-$59.536865 & 14.63   & ~ & ~ & 1.0 & 20.10     \\ 
        Star109 & 5350364185134713600 & 160.943094 & $-$59.520524 & 14.63   & 2CXO J104346.3-593113 & 1.157e$-$13 & ~ &  ~  \\ 
        Star110 & 5350363154342558720 & 161.034359 & $-$59.560266 & 14.68   & ~ & ~ & ~ &  ~  \\ 
        Star112 & 5350363635378755328 & 160.913461 & $-$59.567964 & 14.69   & 2CXO J104339.2-593404 & 4.735e$-$15 & ~ &  ~  \\ 
        Star113 & 5350387206160728832 & 161.066756 & $-$59.493619 & 14.73   & ~ & ~ & ~ &  ~  \\ 
        Star114 & 5350362776370302592 & 161.056217 & $-$59.597553 & 14.73   & 2CXO J104413.6-593550 &   ~& ~ &  ~  \\ 
        Star116 & 5350363635378740480 & 160.907099 & $-$59.572874 & 14.76   & ~ & ~ & ~ &  ~  \\ 
        Star117 & 5350376039244557952 & 160.899717 & $-$59.498357 & 14.82   & 2CXO J104335.9-592954 & 1.369e$-$15 & ~ &  ~  \\ 
        Star118 & 5350375970525073152 & 160.860781 & $-$59.507482 & 14.82   & ~ & ~ & ~ &  ~  \\ 
        Star120 & 5350363188702319744 & 161.018859 & $-$59.543911 & 14.87   & 2CXO J104404.5-593238 & 2.798e$-$14 & ~ &  ~  \\ 
        Star121 & 5350363051263288832 & 161.014753 & $-$59.573916 & 14.89   & ~ & ~ & ~ &  ~  \\ 
        Star123 & 5350386931282829568 & 161.104021 & $-$59.504402 & 15.15   & ~ & ~ & 1.0 & 157.30     \\ 
        Star124 & 5350362845104924672 & 161.10796 & $-$59.586224 & 15.15   & ~ & ~ & ~ &  ~  \\ 
        Star126 & 5350363429220322560 & 160.926396 & $-$59.574656 & 15.21   & 2CXO J104342.3-593428 & 1.12e$-$14 & ~ &  ~  \\ 
        Star127 & 5350362467147580416 & 160.984236 & $-$59.632169 & 15.38   & ~ & ~ & ~ &  ~  \\ 
        Star128 & 5350363326141055360 & 160.93641 & $-$59.603614 & 15.46   & ~ & ~ & ~ &  ~  \\ 
        Star130 & 5350363429220314752 & 160.926542 & $-$59.578734 & 15.54   & ~ & ~ & ~ &  ~  \\ 
        Star131 & 5350363223061830656 & 160.961495 & $-$59.61846 & 15.55   & ~ & ~ & ~ &  ~  \\ 
        Star132 & 5350387790276226432 & 160.947971 & $-$59.478465 & 15.58   & ~ & ~ & ~ &  ~  \\ 
        Star133 & 5350363841537247872 & 160.941632 & $-$59.55154 & 15.62   & 2CXO J104346.0-593305 & 7.592e$-$16 & 2.0 & 10.40 $\pm$ 10.00 \\ 
        Star135 & 5350363669738521600 & 160.906405 & $-$59.551207 & 15.72   & ~ & ~ & ~ &  ~  \\ 
        Star136 & 5350386931282819584 & 161.117175 & $-$59.517364 & 15.75   & ~ & ~ & ~ &  ~  \\ 
        Star138 & 5350386553325676800 & 161.095694 & $-$59.521801 & 15.87   & ~ & ~ & ~ &  ~  \\  
        \\
        \hline
    \end{tabular}
\end{table*}


\bibliography{references}{}
\bibliographystyle{aasjournal}

\end{document}